\documentclass[manuscript]{acmart}

\usepackage{graphicx}
\usepackage{subcaption}

\AtBeginDocument{%
  }

\setcopyright{acmlicensed}
\copyrightyear{2018}
\acmYear{2018}
\acmDOI{XXXXXXX.XXXXXXX}

\acmConference[Conference acronym 'XX]{Make sure to enter the correct
  conference title from your rights confirmation emai}{June 03--05,
  2018}{Woodstock, NY}
\acmISBN{978-1-4503-XXXX-X/18/06}

\begin{document}

\title[EchoSim4D: XR Echocardiography Training Simulator]{EchoSim4D: A Proof-of-Concept Gamified XR Echocardiography Training Simulator for Neonates using 4D Ultrasound Volume}

\author{Deepthy Rose Jose}
\email{am20s052@smail.iitm.ac.in}
\orcid{0009-0005-6632-7596 }
\affiliation{%
  \institution{Indian Institute of Technology Madras}
  \city{Chennai}
  \state{Tamil Nadu}
  \country{India}
}

\author{Venkataseshan Sundaram}
\email{venkatpgi@gmail.com}
\orcid{0000-0002-3135-8115}
\affiliation{%
  \institution{Post Graduate Institute of Medical Education \& Research}
  \state{Chandigarh}
  \country{India}
}

\author{M Manivannan}
\email{mani@iitm.ac.in}
\orcid{0000-0003-1162-1550}
\affiliation{%
  \institution{Indian Institute of Technology Madras}
  \city{Chennai}
  \state{Tamil Nadu}
  \country{India}
}

\renewcommand{\shortauthors}{Jose et al.}

\begin{abstract}
Neonatal echocardiography is vital for early detection of heart anomalies in newborns, enabling timely, non-invasive interventions where 4D ultrasound, adds the dimension of time to 3D imaging, enhances diagnostic capabilities by visualizing real-time heart dynamics. However, training for 4D neonatal echocardiography is limited by the lack of simulators that support 4D Ultrasound volume visualization within gamified environments.

This paper introduces EchoSim4D, an XR-based simulator leveraging novel pipeline for visualizing 4D volume data in Unity, incorporating real-time volume reconstruction, and a preloaded version optimized for low-end systems. EchoSim4D integrates a sensor-equipped manikin and a custom 3D-printed transducer with a 6-DOF sensor, replicating the precise probe maneuvers necessary for neonatal echocardiography. In a validation study with postgraduate medical students (0-5 years of experience), supervised by a domain expert, EchoSim4D demonstrated high visual fidelity and training efficacy. Findings suggest that 4D visualization techniques hold significant potential for advancing medical training in neonatal echocardiography.
\end{abstract}



\begin{teaserfigure}
  \includegraphics[width=\textwidth]{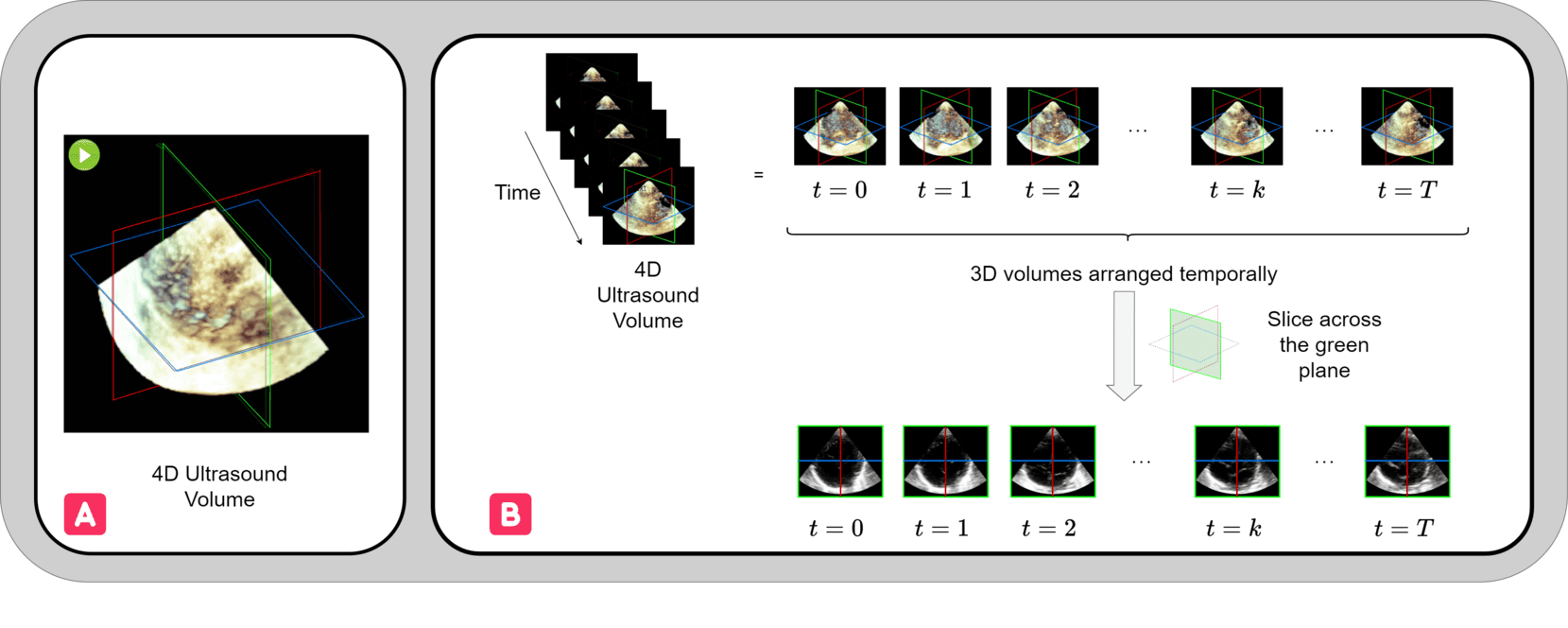}
  \caption{a) 4D ultrasound volume obtained using Philips X7-2 Ultrasound probe. b) Slicing workflow for viewing the axes of interest in a 4D volume.}
  \Description{}
  \label{fig:teaser}
\end{teaserfigure}

\received{20 February 2007}
\received[revised]{12 March 2009}
\received[accepted]{5 June 2009}

\maketitle

\section{Introduction}
Neonatal echocardiography is a critical diagnostic tool that requires significant skill and precision, particularly in handling the probe and interpreting real-time ultrasound data. Despite its importance, training opportunities for neonatal echocardiography are limited, as they demand extensive practice \cite{psychomotor} and access to specialized equipment. Existing training simulators primarily rely on (2D + Time) ultrasound images and lack the immersive, hands-on experience necessary for developing proficiency in the required probe maneuvers and anatomical identification.

To address these limitations, we present EchoSim4D, an innovative proof-of-concept XR echocardiography training simulator designed specifically for neonatal applications. EchoSim4D leverages 4D volumetric data within Unity \cite{Unity}, creating an immersive environment where trainees can practice essential probe maneuvers, such as tilt and rotation, while interacting with a custom-designed 3D-printed probe and sensor-equipped manikin. This simulator bridges the gap between theoretical learning and practical application, providing an accessible, gamified approach to training.

EchoSim4D not only facilitates real-time ultrasound data visualization but also incorporates a tiered gamification structure to enhance engagement and learning retention. The system's architecture allows for smooth transitions between different anatomical views, while a built-in feedback mechanism guides users through their tasks. By integrating visual fidelity and real-time feedback, EchoSim4D aims to simulate the authentic experience of neonatal echocardiography and improve trainees' diagnostic skills.

This paper details the development and evaluation of EchoSim4D, with contributions that include a novel pipeline for 4D ultrasound data visualization in a game engine, the design and implementation of the simulator, and insights from a validation study with postgraduate medical students. We demonstrate how EchoSim4D advances echocardiography training by offering a realistic, scalable, and cost-effective solution tailored for neonatal care.

\section{Background and Related Work}

\begin{figure}[h]
  \centering
  \begin{subfigure}{0.15\textwidth}
    \centering
    \includegraphics[width=\linewidth]{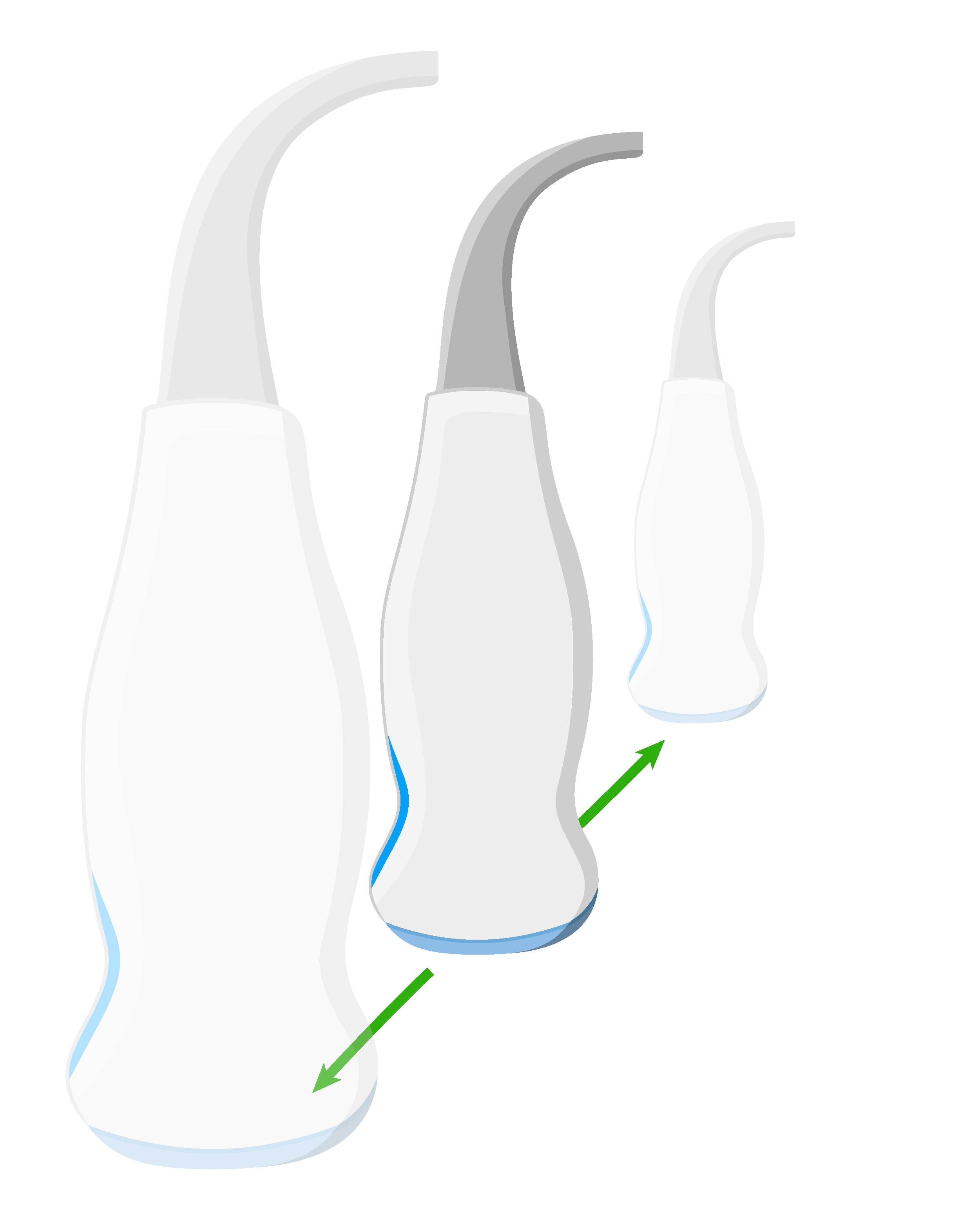}
    \caption{}
    \label{fig:slide}
  \end{subfigure}
  \hfill
  \begin{subfigure}{0.15\textwidth}
    \centering
    \includegraphics[width=\linewidth]{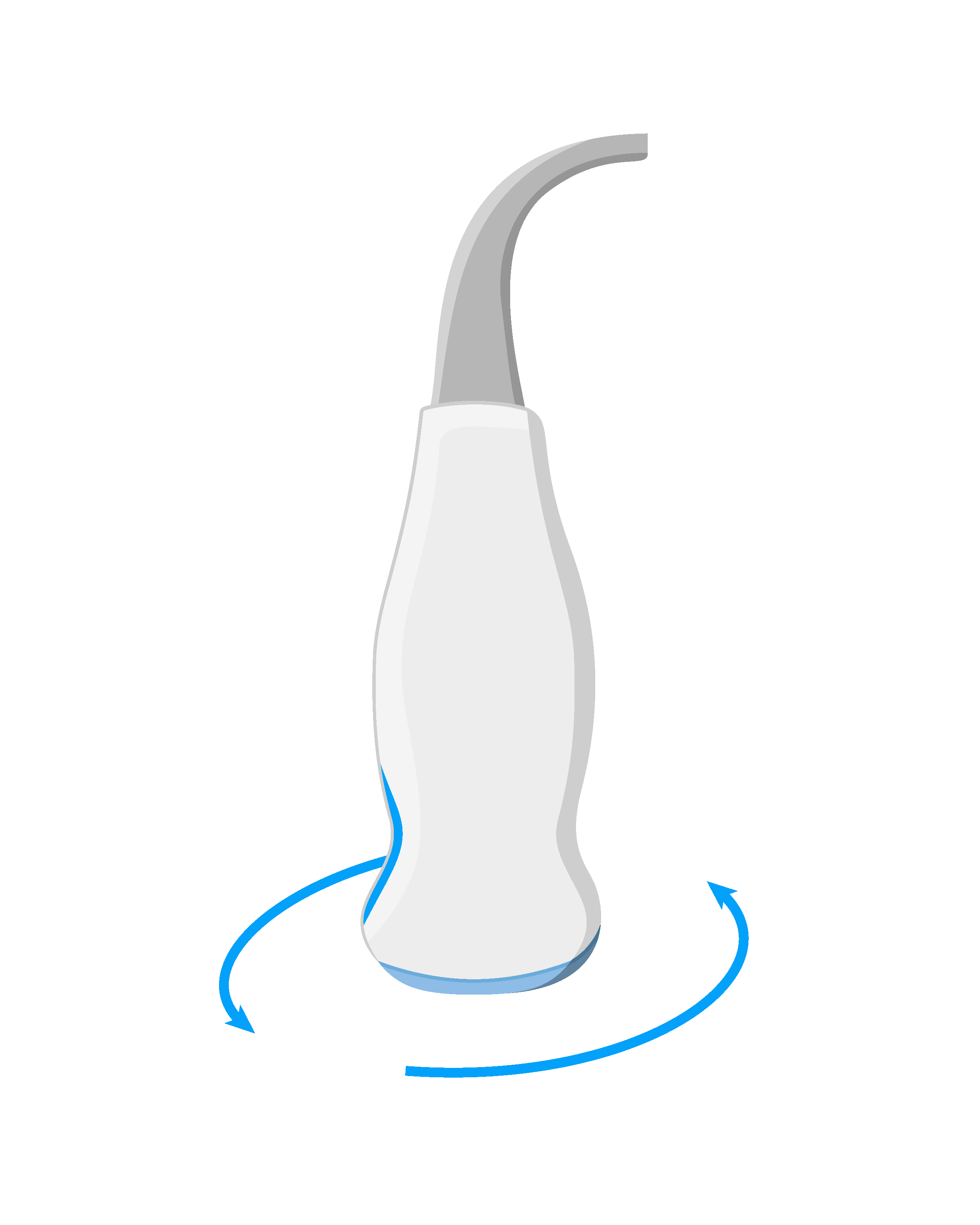}
    \caption{}
    \label{fig:rotate}
  \end{subfigure}
  \hfill
  \begin{subfigure}{0.15\textwidth}
    \centering
    \includegraphics[width=\linewidth]{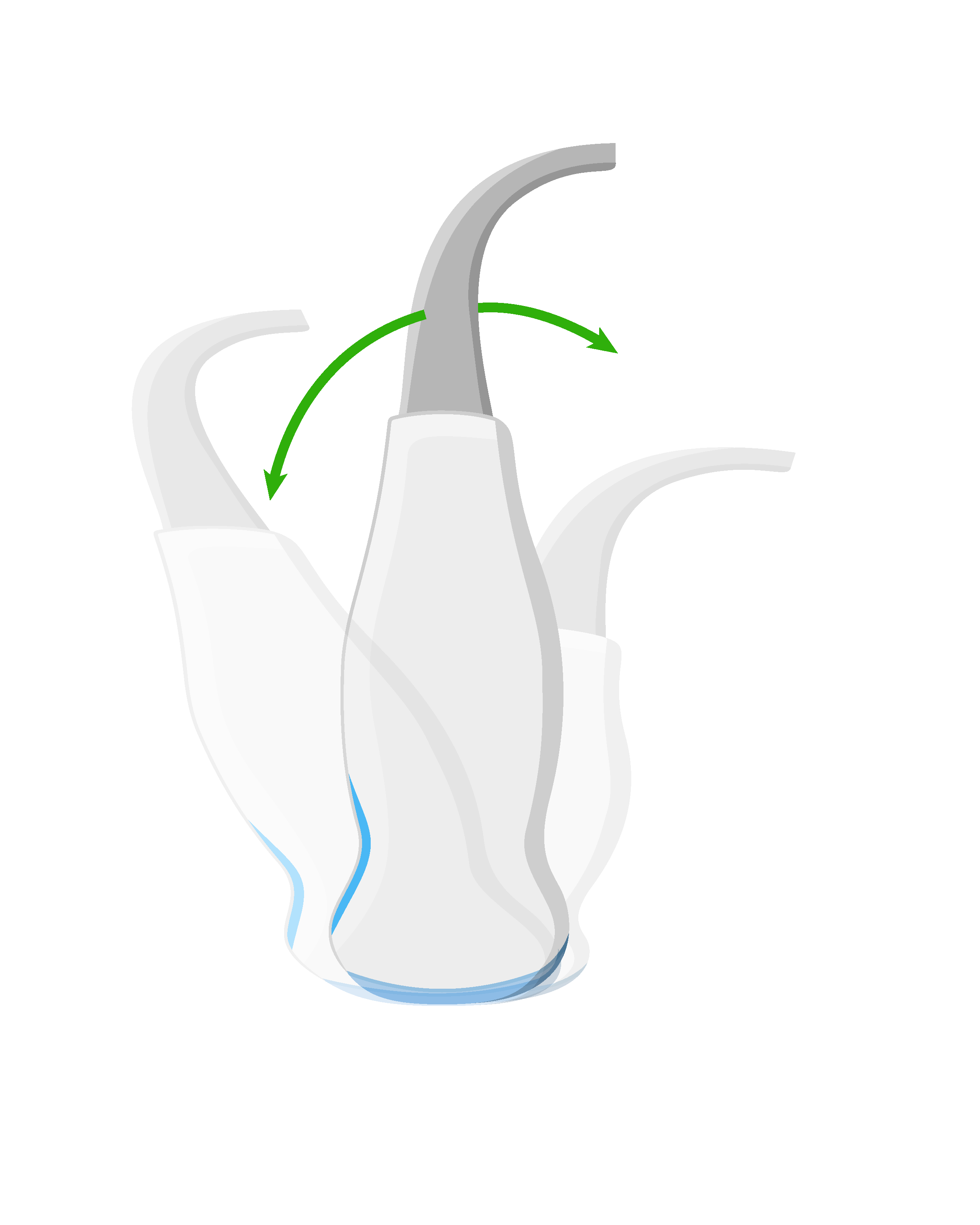}
    \caption{}
    \label{fig:tilt}
  \end{subfigure}
  \hfill
  \begin{subfigure}{0.15\textwidth}
    \centering
    \includegraphics[width=\linewidth]{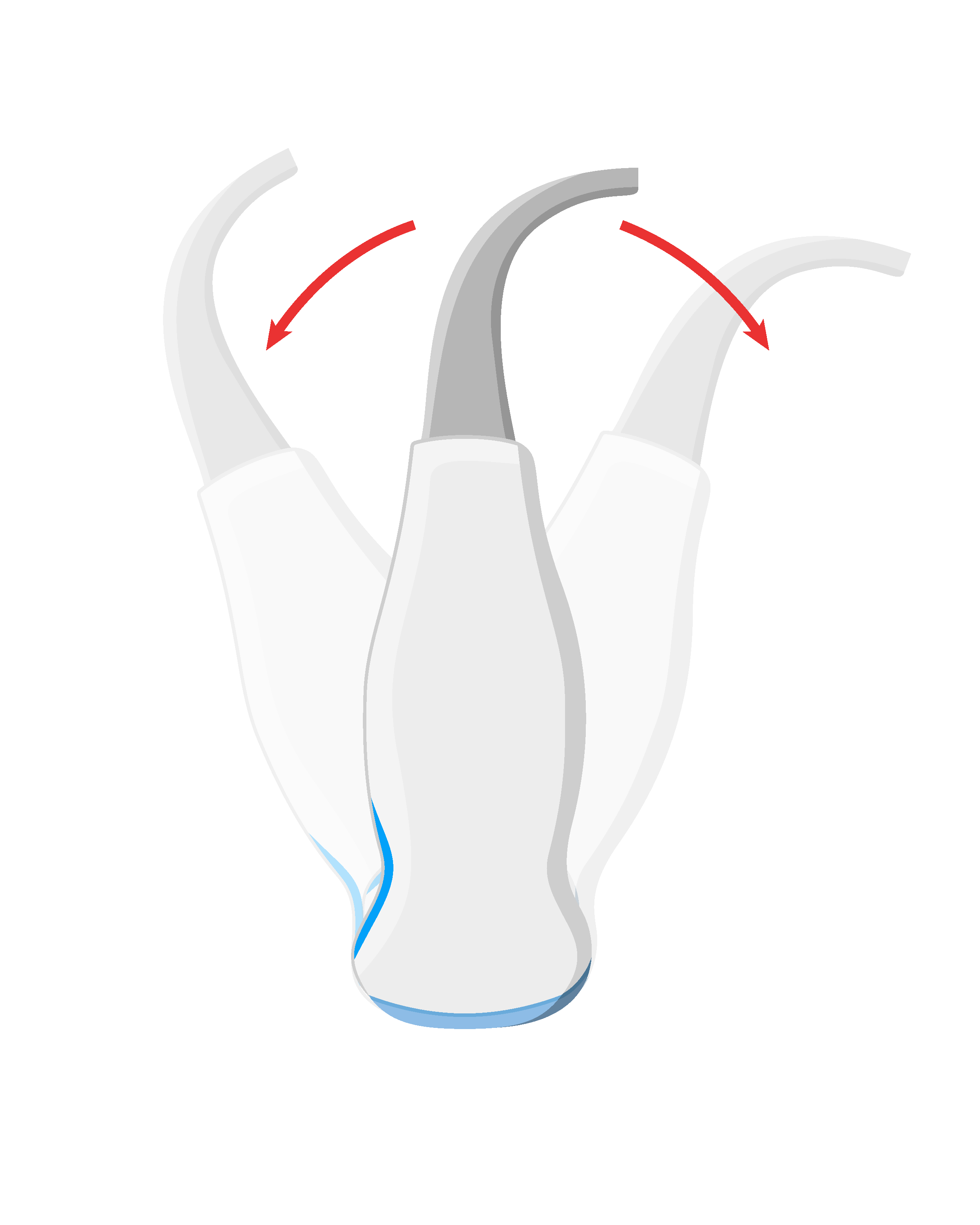}
    \caption{}
    \label{fig:rock}
  \end{subfigure}
  \hfill
  \begin{subfigure}{0.15\textwidth}
    \centering
    \includegraphics[width=\linewidth]{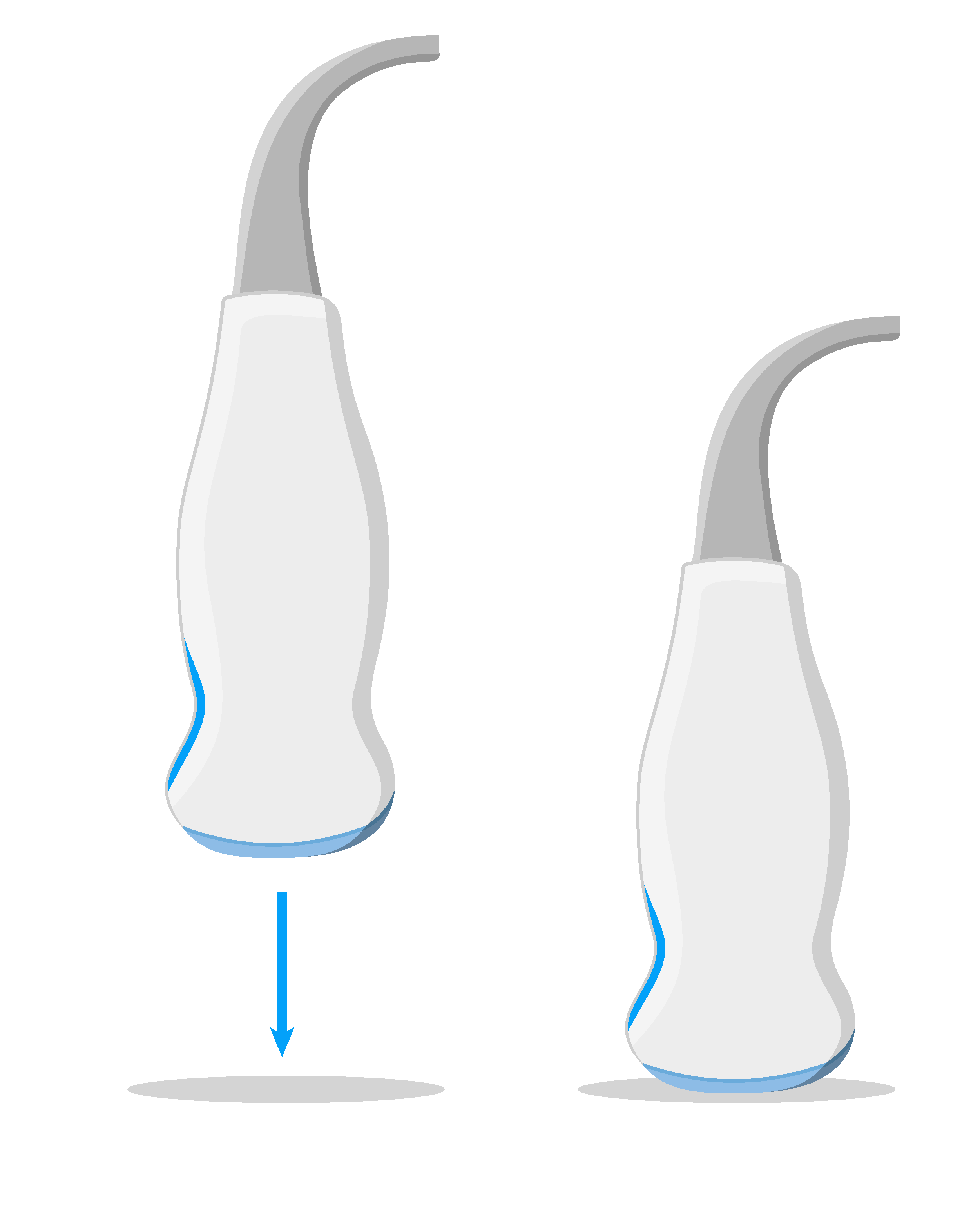}
    \caption{}
    \label{fig:pressure}
  \end{subfigure}
  \hfill
  \begin{subfigure}{0.15\textwidth}
    \centering
    \includegraphics[width=\linewidth]{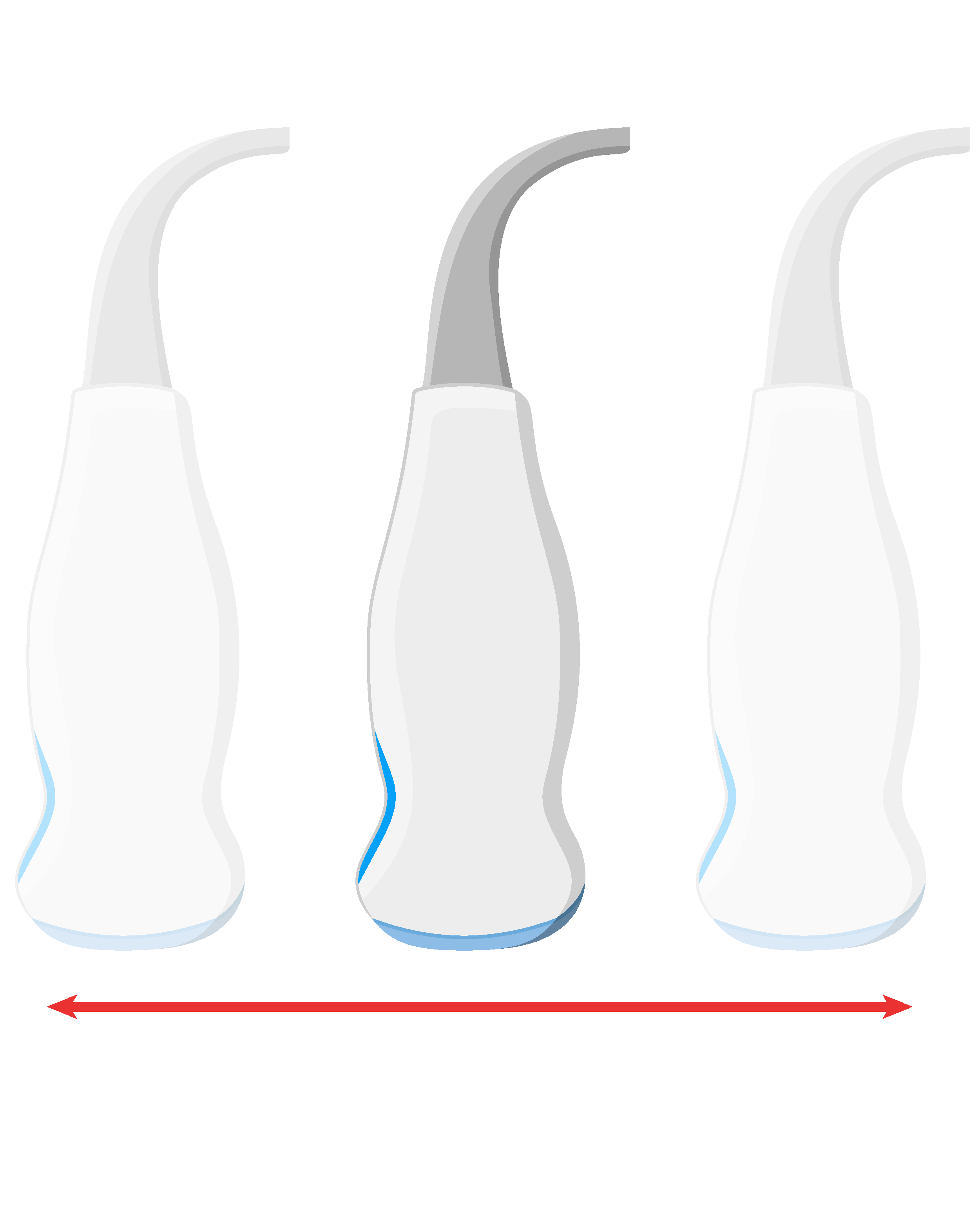}
    \caption{}
    \label{fig:sweep}
  \end{subfigure}
  
  \caption{\textbf{Standard Probe Maneuvers}: a) \textit{Sliding}, b) \textit{Rotating}, c) \textit{Tilting}, d) \textit{Rocking}, e) \textit{Pressure}, and f) \textit{Sweeping} }
  \Description{A single row of six images, each with its own caption.}
  \label{fig:maneuver}
\end{figure}

Advancements in medical simulators have significantly enhanced training environments. However, traditional echocardiography tools, primarily focused on 2D imaging, fall short in replicating the complexities of neonatal procedures, which require nuanced probe handling and 3D visualization. Existing platforms like HeartWorks \cite{heartworks} and CAE Vimedix \cite{vimedix} cater to adult cardiology but lack specific tools for neonatal anatomy.
To improve anatomical accuracy, recent studies have explored 3D and virtual reality (VR)-based simulators. Platforms such as SonoSim \cite{sonosim} provide enhanced 3D views but do not fully support real-time feedback or meet the specific demands of neonatal echocardiography. Game engines like Unity \cite{Unity} and Unreal Engine \cite{unrealengine} are increasingly utilized in medical simulations, integrating real-time 3D ultrasound. Despite these advancements, 4D visualization—which incorporates time as a fourth dimension into 3D imaging—remains underexplored due to high computational demands.
4D ultrasound (Figure \ref{fig:teaser}a) is an advanced imaging technique that extends three-dimensional (3D) ultrasound (Figure \ref{fig:teaser}b) by incorporating the dimension of time, enabling real-time visualization of dynamic physiological processes. This modality captures volumetric data in the form of voxels, three-dimensional equivalents of pixels, that encode spatial information along the length, width, and depth axes. The resulting comprehensive cross-sectional views facilitate detailed spatial understanding, which is crucial in complex clinical contexts such as cardiac imaging. Unlike traditional 2D+time ultrasound that captures time-sequenced slices lacking depth, 4D ultrasound provides a complete three-dimensional view over time.
Interpreting 4D ultrasound data in neonatal cardiac imaging is complex and requires specialized training modules. These modules equip healthcare professionals with the skills to accurately interpret 4D data and handle probes effectively, reducing the risk of misdiagnosis in this vulnerable population. The mastery of ultrasound probe maneuvers (Figure \ref{fig:maneuver}) -such as sliding, rotation, tilting, angulation, and sweeping - is crucial. The clock system standardizes the orientation of the probe, enabling consistent image acquisition and interpretation.

The DICOM standard is essential for storing and transmitting ultrasound data, including critical metadata. However, inconsistencies in 4D DICOM encoding among manufacturers like Philips, GE, and Siemens hinder the development of interoperable training simulators. Tailored visualization solutions that accommodate these proprietary formats can enhance compatibility and support versatile training modules across different systems.

Gamification has proven effective in medical training \cite{gamificliver} \cite{gamificationmeded} by boosting engagement through elements like levels and feedback. Nevertheless, ultrasound simulators utilizing 4D data in a gamified format are rare. Platforms like Unity enable the development of interactive simulators that integrate 4D data visualization and gamified elements, providing a controlled environment for skill acquisition. This approach is especially valuable in neonatal echocardiography due to the risks of practicing on real patients. Gamified training simulators improve user engagement and facilitate exploratory learning, allowing trainees to build practical skills and confidence safely.

EchoSim4D aims to bridge these gaps by combining 4D volume visualization with gamified, tiered tasks tailored to neonatal echocardiography. It provides real-time feedback and enhanced anatomical immersion, addressing current limitations and offering a novel tool for skill development in neonatal ultrasound training.

Ultimately, the development of a robust 4D visualization technique for neonatal cardiac ultrasound can have far-reaching implications. While this research focuses on a specific use case, the insights gained can serve as a foundation for broader advancements in medical imaging. By addressing the technical challenges of 4D data visualization and DICOM compatibility, this work opens up the possibility of creating more effective training simulators across various medical fields, thereby improving both education and patient outcomes.

\section{Case Study for Neonatal Echocardiography: Transthoracic Echocardiography (TTE)}

\subsection{Transthoracic Echocardiography (TTE)}

Transthoracic Echocardiography (TTE) was chosen as a case study for a neonatal echocardiography simulator because it is the primary, non-invasive method used for cardiac assessment in newborns. TTE is safe, provides essential diagnostic insights, and is widely used in clinical practice, making it ideal for training. Its portability allows for bedside use, and real-time imaging helps develop skills in live analysis, which is crucial in neonatal care. By simulating TTE, trainees can gain practical experience with the standard tool for diagnosing and monitoring neonatal heart conditions.

\subsection{Core Expertise Required for TTE}

\begin{figure}[h]
  \centering
  \begin{minipage}{0.9\textwidth}
    \centering
    \captionof{table}{\textbf{Standard Transthoracic Echocardiographic Views}}
    \label{tab:views}
    \begin{tabular}{ccl}
      \toprule
      View & Probe Notch Direction & Tilt Range (in degrees) \\
      \midrule
      Apical & 3 o'clock & 5 to 10 \\
      PLAX & 11 o'clock & 5 to 10 \\
      PSAX & 1 o'clock & 5 to 10 \\
      Subcostal & 3 o'clock & 40 to 45 \\
      Suprasternal & 1 o'clock & 5 to 10 \\
      \bottomrule
    \end{tabular}
  \end{minipage}

  \vspace{1em} 

  \begin{minipage}{0.45\textwidth}
    \centering
    \includegraphics[width=\textwidth]{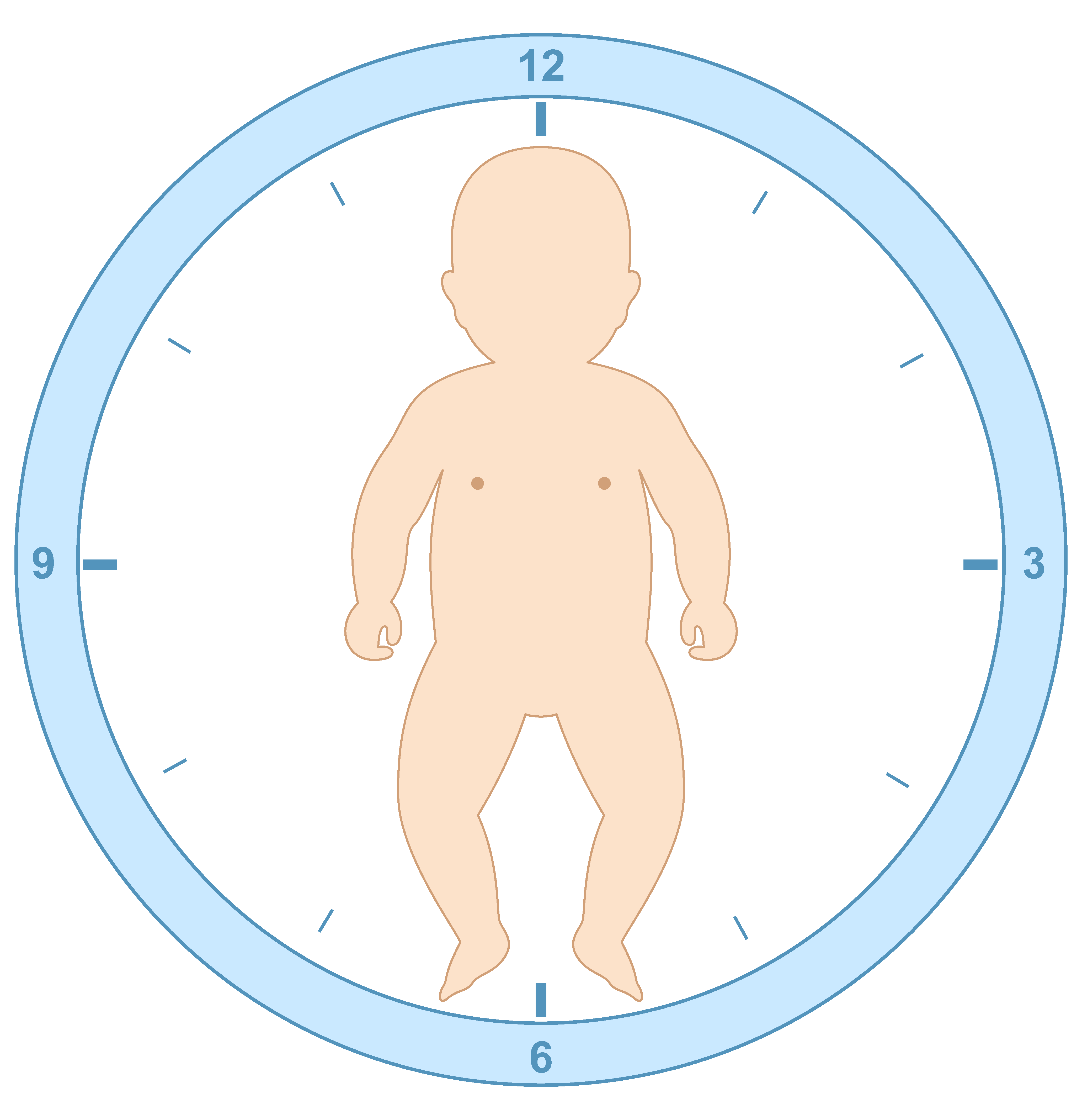}
    \caption{Clock System}
    \Description{Some description}
  \end{minipage}%
  \hfill%
  \begin{minipage}{0.45\textwidth}
    \centering
    \includegraphics[width=\textwidth]{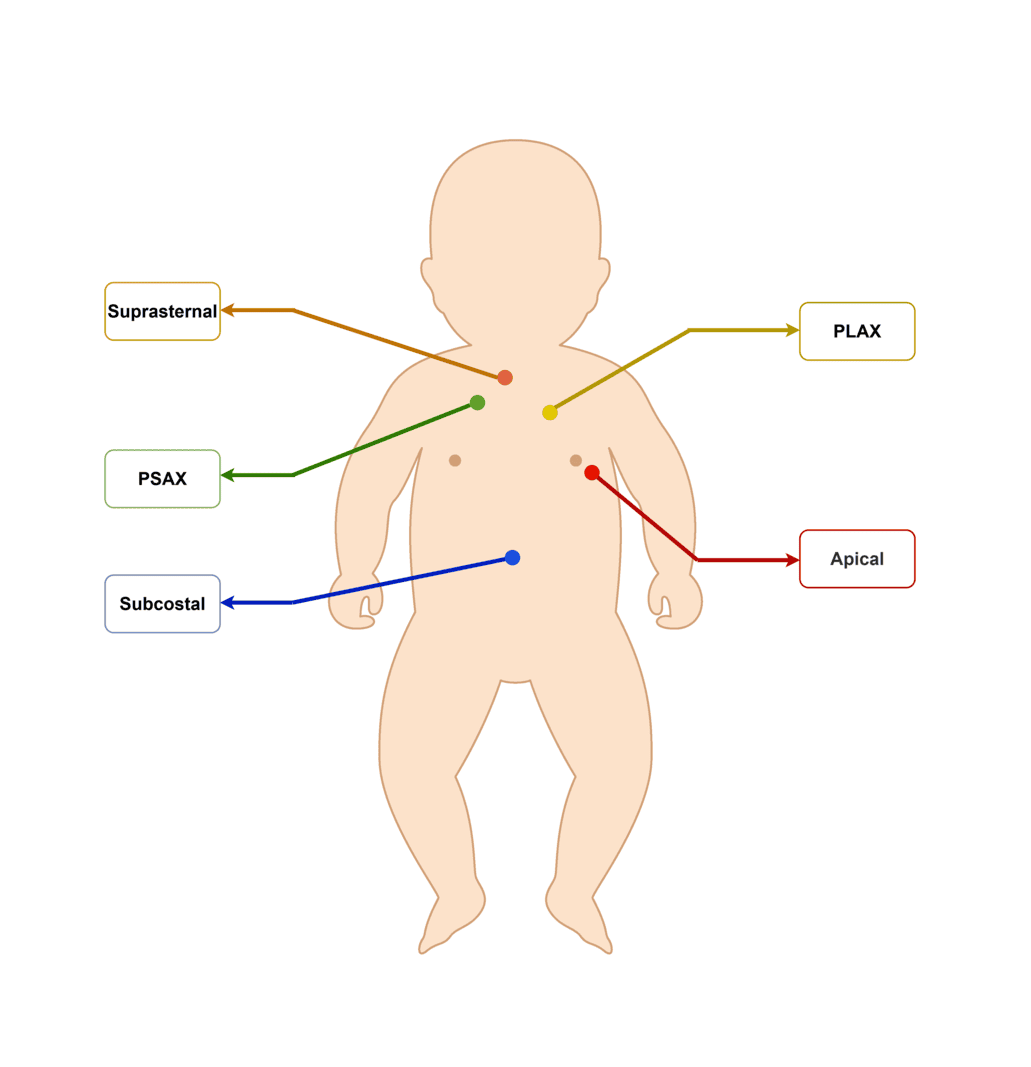}
    \caption{Standard TTE Locations}
    \Description{Some description}
  \end{minipage}
\end{figure}

Mastery in neonatal echocardiography TTE necessitates proficiency in key views, including Apical, Parasternal Long-Axis (PLAX), Parasternal Short-Axis (PSAX), Subcostal, and Suprasternal views (Table \ref{tab:views}) These views are fundamental for diagnosing various cardiac anomalies, such as congenital heart defects and pulmonary hypertension. Proficient probe manipulation across anatomical planes and knowledge of correct clock angles and tilt ranges are vital to obtaining diagnostically valuable images.
In standard echocardiographic practice, normal views are achieved by aligning the probe to standard anatomical positions, yielding essential cross-sectional images. Tilt views enhance visualization of structures less visible in normal views by adjusting the probe’s angle. EchoSim4D incorporates the following views to simulate neonatal cardiac assessment: Apical (Normal, Tilt), PSAX (Normal, Tilt), PLAX (Normal, Tilt), Subcostal (Normal, Tilt), and Suprasternal (Normal, Tilt).

\subsection{Learning Objectives for Training}
Upon completion of the training, users should be able to:
1.	\textbf{Identify Target Chamber Locations}: Accurately locate standard TTE windows for neonatal echocardiography.
2.	\textbf{Orient and Angle for Normal Views}: Properly orient the probe notch according to the clock system  and understand the optimal angle range for normal views. For instance, the Apical view generally involves positioning the probe at the chest apex at a 3 o'clock position.
3.	\textbf{Optimize Tilt Angles for Enhanced Visualization}: Understand how to tilt the probe to achieve optimal views of specific anatomical structures. This includes knowing the tilt angle ranges needed to visualize key structures, such as the aortic valve in the Apical view or ventricular outflow tracts in the PSAX view.

\section{Design Process}
\subsection{Data Acquisition}

\begin{figure}[h]
  \centering
  \begin{subfigure}{0.2\textwidth}
    \centering
    \includegraphics[height=4cm]{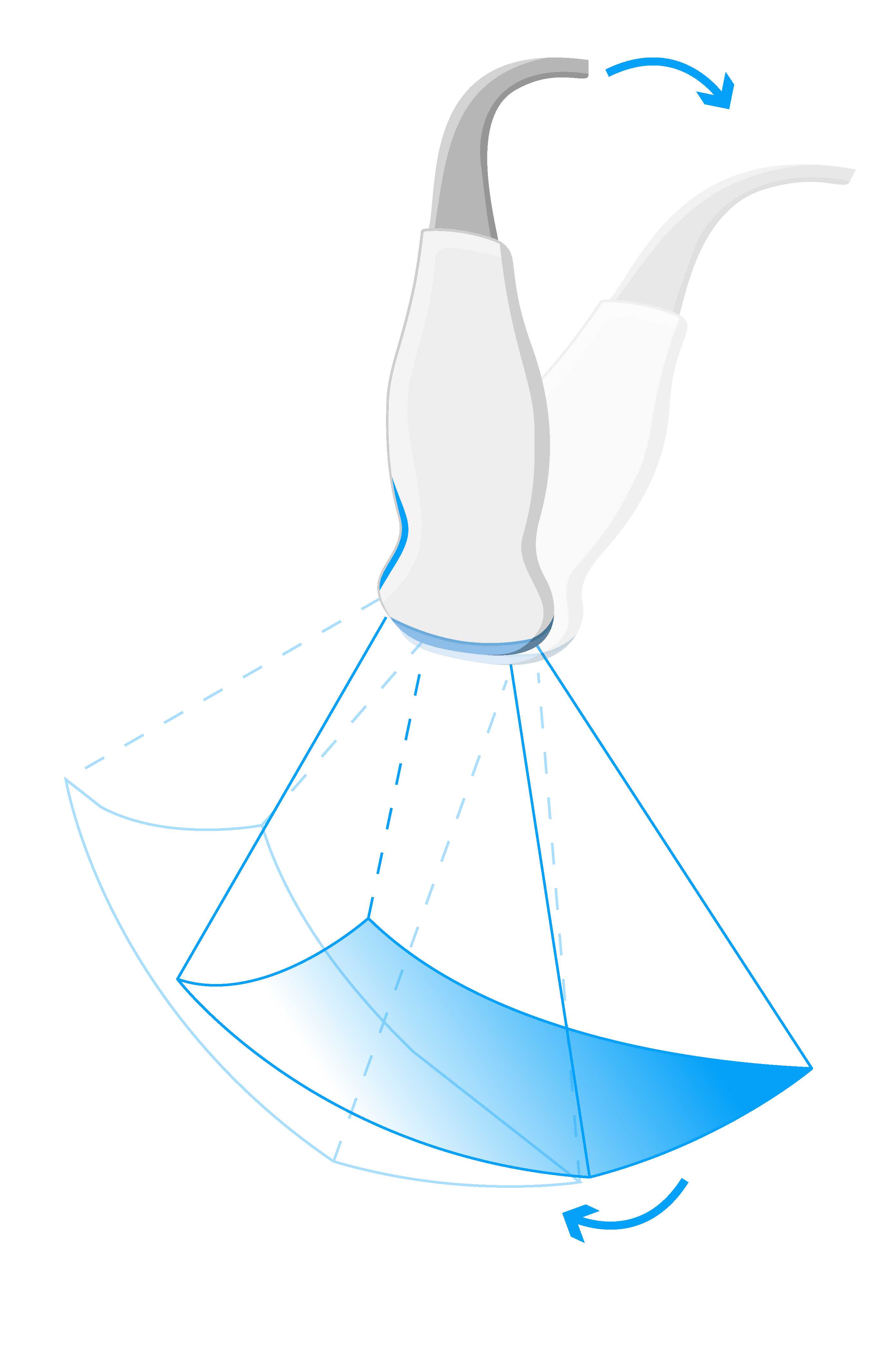}
    \caption{}
    \label{fig:4dcap}
  \end{subfigure}
  \hspace{0.5cm}
  \begin{subfigure}{0.7\textwidth}
    \centering
    \includegraphics[height=4cm]{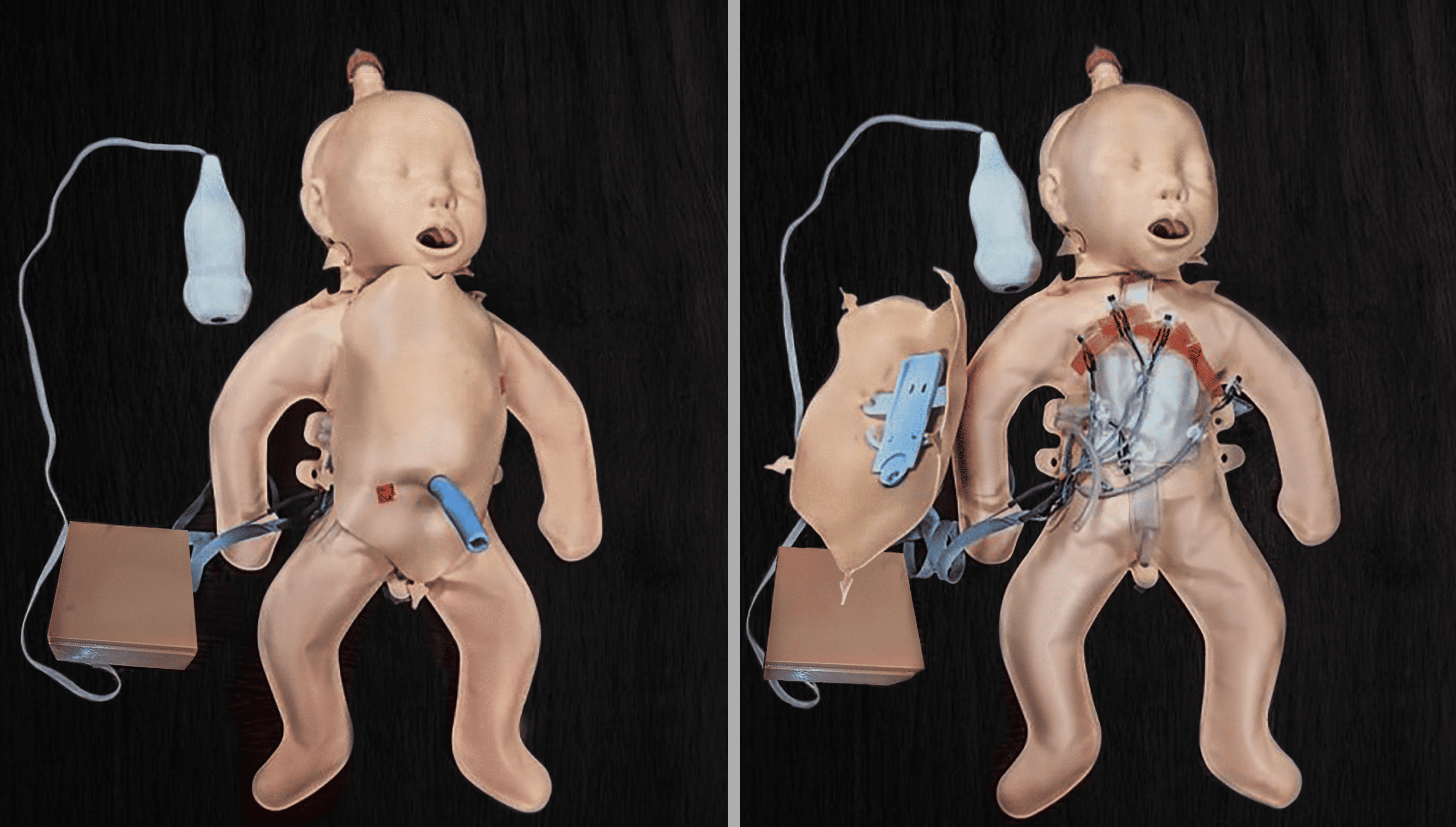}
    \caption{}
    \label{fig:sim_heard}
  \end{subfigure}
  
  \caption{a) \textit{4D Volume Data Acquisition}, b) \textit{EchoSim4D Hardware}}
\end{figure}

Echocardiographic data were acquired using a Philips EPIQ 7 ultrasound system with a Philips X7-2 matrix transducer. For each neonate, up to ten 4D DICOM volumes (3D + time) were collected from five standard transthoracic echocardiographic views: Apical, Parasternal Short Axis (PSAX), Parasternal Long Axis (PLAX), Subcostal, and Suprasternal, covering both standard and tilt angle perspectives. All images were recorded in DICOM format, ensuring compliance with clinical standards, and were acquired based on clinical indications assessed by the attending neonatologists. The X7-2 transducer can produce volumetric datasets that, depending on the quality and duration of acquisition, may even reach gigabyte-scale storage requirements.
The transducer replicates the precise probe maneuvers required to identify standard TTE locations and accurately captures the 4D data (Figure \ref{fig:4dcap}).
To optimize these datasets for visualization within the Unity game engine, it was necessary to reduce file size while preserving crucial anatomical details. The following strategies were considered during the data acquisition phase:
\begin{enumerate}
    \item \textit{Reducing Beat Count}: The standard “3D beats 4” protocol, which captures four cardiac cycles for enhanced precision, was modified to the “3D beats 1” protocol directly on the EPIQ 7 system. This adjustment, capturing a single cardiac cycle, significantly decreased the data volume while maintaining sufficient anatomical fidelity for training. As a result, the ultrasound volumes were reduced to approximately 20 MB per dataset as an example, considerably smaller than the sizes associated with multi-beat acquisitions.
    \item \textit{Consideration of Acquisition Time}: Although reducing acquisition time could further lower data size, this method was inherently linked to the beat count reduction and was not prioritized independently.
    \item \textit{Sector/Wedge Size Reduction}: While reducing the sector size could further minimize file size, this approach was not adopted as it could lead to the loss of critical anatomical details necessary for accurate training.
\end{enumerate}

By reducing the acquisition to a single beat, substantial data size reduction was achieved, resulting in 4D ultrasound volumes that were, for example, around 20 MB each. Each dataset was subsequently reviewed by a domain expert to confirm the presence of essential anatomical transitions, ensuring suitability for training before additional preprocessing for Unity visualization.

\subsection{Data Preprocessing}

\begin{figure}[h]
  \centering
  \includegraphics[width=0.8\textwidth]{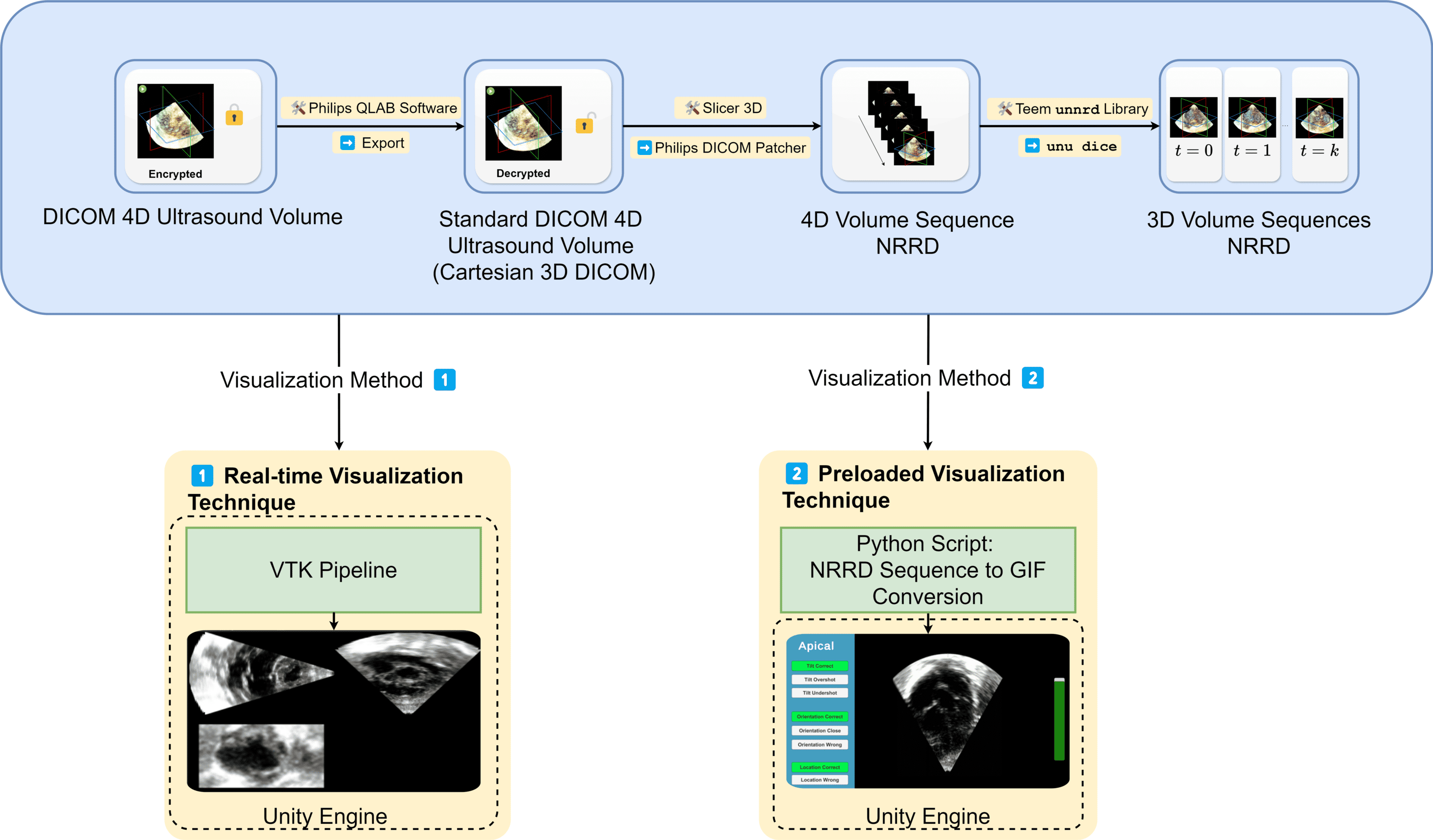}
  \caption{Data Preprocessing Pipeline}
  \label{fig:data_pp}
\end{figure}

\begin{figure}[h]
  \centering
  \includegraphics[width=0.8\textwidth]{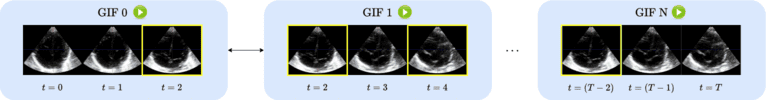}
  \caption{\textbf{Preloaded Visualization}: Configuration for Extracting Multiple GIFs}
  \label{fig:gif_conv}
\end{figure}

The acquired ultrasound data was converted into a Unity-compatible format using the data preprocessing pipeline illustrated in (Figure \ref{fig:data_pp}). This pipeline facilitates the transformation of DICOM 4D ultrasound data into a format suitable for real-time and preloaded visualizations in Unity. The process involves several steps:
\begin{enumerate}
    \item \textbf{Data Decryption and Export}: The encrypted 4D DICOM volume is decrypted using Philips QLAB software and exported to a standard Cartesian 4D DICOM format.
    \item \textbf{Conversion to NRRD Format}: Using Slicer 3D along with the Philips DICOM Patcher, the 4D DICOM data is converted into a 4D NRRD (Nearly Raw Raster Data) sequence. This conversion allows compatibility with open-source libraries.
    \item \textbf{Volume Sequence Extraction}: The Teem unrrd Library extracts 3D volume sequences from the 4D NRRD file, enabling frame-by-frame processing of the ultrasound data.
    \item \textbf{Visualization Techniques}: \textit{Real-time Visualization} (Figure \ref{fig:data_pp}): Using the Visualization Toolkit (VTK) plugin \cite{vtkBook} for Unity, real-time volumetric rendering is performed. This approach allows for dynamic interactions, such as cutting the volume along arbitrary planes, providing flexibility in visualization. However, it demands significant computational resources and is better suited for high-performance systems due to Unity’s single-threaded architecture, which can lead to resource bottlenecks, especially during cross-platform XR development. Therefore, only the primary imaging planes are visualized, with potential for further optimization within the Unity environment. \textit{Preloaded Visualization} (Figure \ref{fig:gif_conv}): The NRRD sequence is converted into GIF format using Python libraries, as Unity does not natively support GIF playback. By preloading the GIFs in Unity, latency during playback is minimized, supporting smoother visualization. This technique is compatible with lower-end systems, as GIF compression significantly reduces file sizes.
\end{enumerate}

This dual visualization approach provides flexibility across different system capabilities. 
For testing EchoSim4D, the Preloaded Visualization technique was selected due to its stability across various compute environments. This approach ensures consistent performance and reliability, maintaining visual fidelity independently of hardware limitations. By preloading data, the visualization was optimized to provide users with a seamless experience, even on systems with lower computational capacities.

\subsection{System Architecture}
\subsubsection{Hardware}

The primary physical model in this study was an air-inflatable baby manikin, designed to replicate neonatal anatomy for echocardiographic training. Analog Hall effect sensors were strategically positioned at transthoracic echocardiography (TTE) standard anatomical locations to accurately capture probe movements. A customized 3D-printed transducer dummy, modelled after the Philips X7-2 probe, was fabricated to maintain realism and ergonomic fidelity. Embedded within this dummy is an MPU6050 inertial measurement unit (IMU) featuring 6 degrees of freedom (DOF), enabling precise measurement of roll, pitch, and yaw during probe maneuvers.
A 5 mm radius circular magnet was affixed to the base of the dummy transducer, facilitating magnetic field variation detection through the analog Hall effect sensors. These sensors detect changes in the magnetic field caused by probe movements, capturing the nuances of position and orientation. An Arduino Nano microcontroller was selected for its compactness and sufficient processing capability, interfacing with the MPU6050 and the five analog Hall effect sensors. This setup allows for accurate data acquisition and real-time feedback essential for motion tracking. To ensure durability and ease of handling, the Nano board is housed in a custom 3D-printed enclosure.

\subsubsection{Software and Gamification}

\begin{figure}[h]
  \centering
  \includegraphics[width=0.95\textwidth]{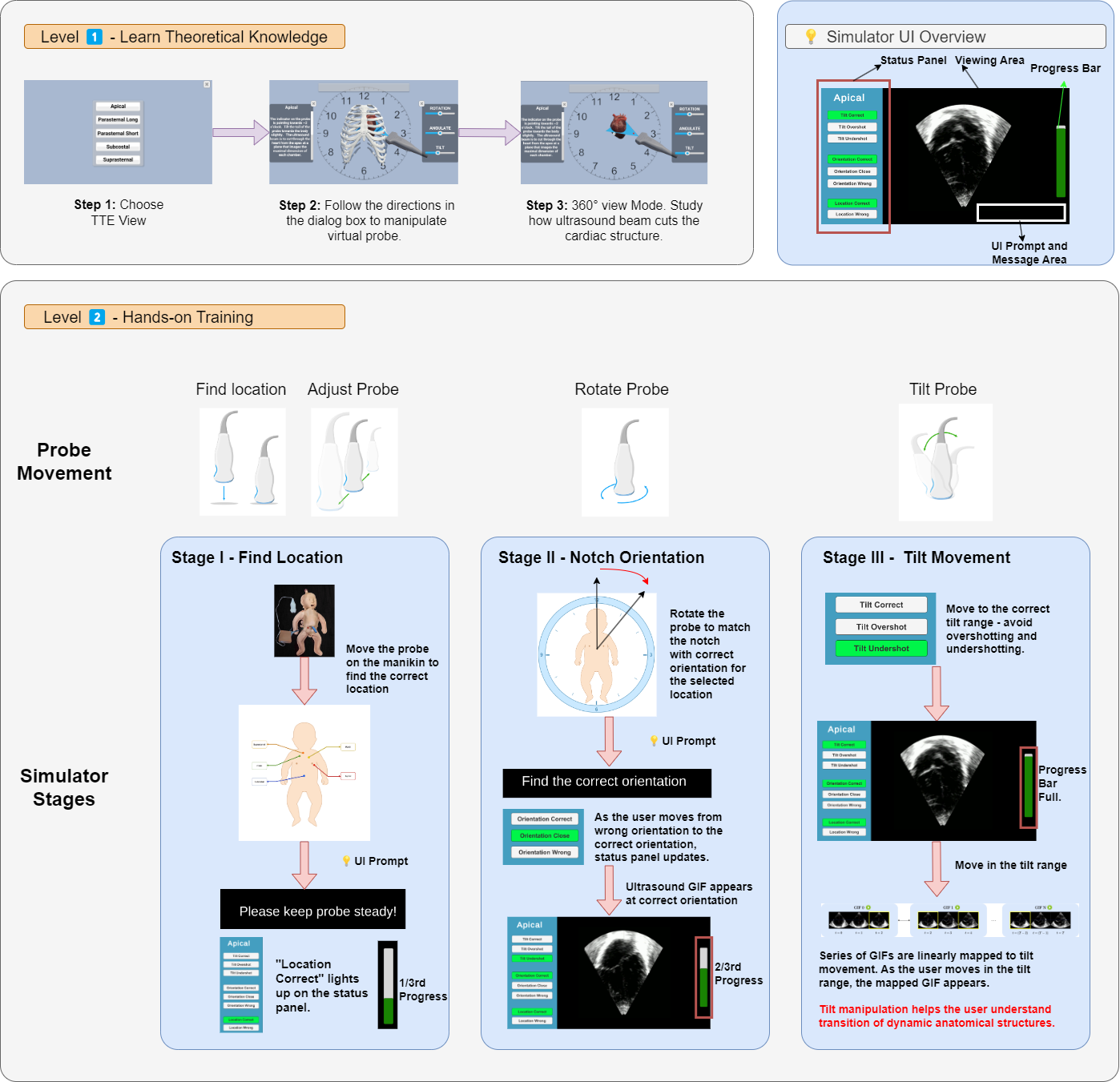}
  \caption{Simulator Flow and Gamification}
  \Description{Some description}
\end{figure}

The EchoSim4D simulator utilizes real-time data parsing of yaw, pitch, roll, and five analog sensor statuses via an Arduino Nano, enabling dynamic rendering within Unity for neonatal echocardiography training. The system is structured into two primary levels:
\begin{itemize}
    \item \textbf{Level 1: Theoretical Knowledge} introduces users to TTE views and probe manipulation fundamentals, laying the groundwork for practical skills.
    \item \textbf{Level 2: Hands-on Training} builds on this foundation, guiding users through locating anatomical regions, aligning the probe notch, and adjusting tilt, with specific tasks for apical view acquisition.
\end{itemize}

Progression is gamified through a status bar indicating:

1/3 Completion for accurate location identification,
2/3 Completion for correct notch alignment, and
3/3 Completion for achieving the correct tilt angle.
The simulator provides sequential feedback during task execution, allowing users to refine their technique. Visualizations in Unity reflect anatomical transitions corresponding to tilt adjustments, facilitating understanding of angle impacts, such as undershot and overshot views. Visualization is contingent on sensor contact and correct orientation, ensuring real-time accuracy and engagement. This tiered approach effectively supports the development of critical skills required for neonatal echocardiography.

\section{Validation Study}

A validation study was conducted to evaluate the efficacy and usability of EchoSim4D, targeting postgraduate medical students with varying experience levels in neonatal echocardiography (0 to 4 years). The study included 10 participants specializing in pediatrics and neonatology, with qualifications ranging from M.D. to D.M. and post-D.M. levels. Under the guidance of a neonatology expert, participants engaged in training scenarios that simulated real-world conditions. M.D. students were classified as novices in neonatal echocardiography due to the limited focus on this technique in their curriculum, while D.M. and post-D.M. students were more familiar with the process.

Participants completed level-based XR training scenarios (\textit{Levels 1 }and \textit{2}) within the simulator, which emphasized aspects of user interaction and skill development. The study used a 44-item questionnaire to assess five core validity dimensions: face, content, construct, ergonomic \& physical, and psychological \& affective validity. Alongside these general dimensions, specific attributes like visual fidelity, gamification, user interaction and feedback, cognitive load, and ergonomic usability were also examined to provide a deeper understanding of EchoSim4D’s unique features.

To evaluate these attributes effectively, a subset of questions from the broader validity questionnaire was utilized. Certain questions naturally spanned multiple dimensions; for instance, aspects of face and content validity played a crucial role in assessing visual fidelity. This integrated approach ensured that each attribute was evaluated in relation to the core validity dimensions, offering a holistic view of EchoSim4D's usability.
Evaluation Metrics
•	Visual Fidelity: Evaluated the simulator's realism, particularly how accurately the ultrasound visualizations align with real-life systems to ensure a true-to-life depiction of cardiac structures.
•	Gamification Effect: Explored the influence of game-like elements, such as levels and tasks, on user engagement and simplifying complex echocardiography techniques.
•	User Interaction \& Feedback: Measured the intuitiveness of physical interactions with the simulator, such as probe handling and feedback mechanisms, contributing to an immersive experience.
•	Cognitive Load: Quantified the mental effort required, reflecting the simulator's balance between complexity and user accessibility.
•	Training Effectiveness: Assessed the simulator's ability to bridge theoretical knowledge with practical echocardiography skills.

\subsection{Results}
The study revealed important insights, highlighting both strengths and areas for improvement. Content validity achieved a mean rating close to 4.0, indicating alignment with neonatal echocardiography principles, while construct validity varied among participants but showed a solid average around 4.0 as well. Ergonomic and physical validity scored an average of 3.7, suggesting variability in user comfort during physical interactions with the simulator.
For usability, visual fidelity scored highly with an average of approximately 4.0, underscoring the realism of the visuals. Cognitive load scores ranged below 3.5 for some users, with an overall average around 3.78, indicating that certain tasks were mentally challenging. Ergonomic usability received an average score of 3.8, mirroring the need for improvements in this area, as seen in the lower scores for ergonomic physical validity.

\begin{figure}[h]
  \centering
  \begin{subfigure}{0.49\textwidth}
    \centering
    \includegraphics[width=\linewidth]{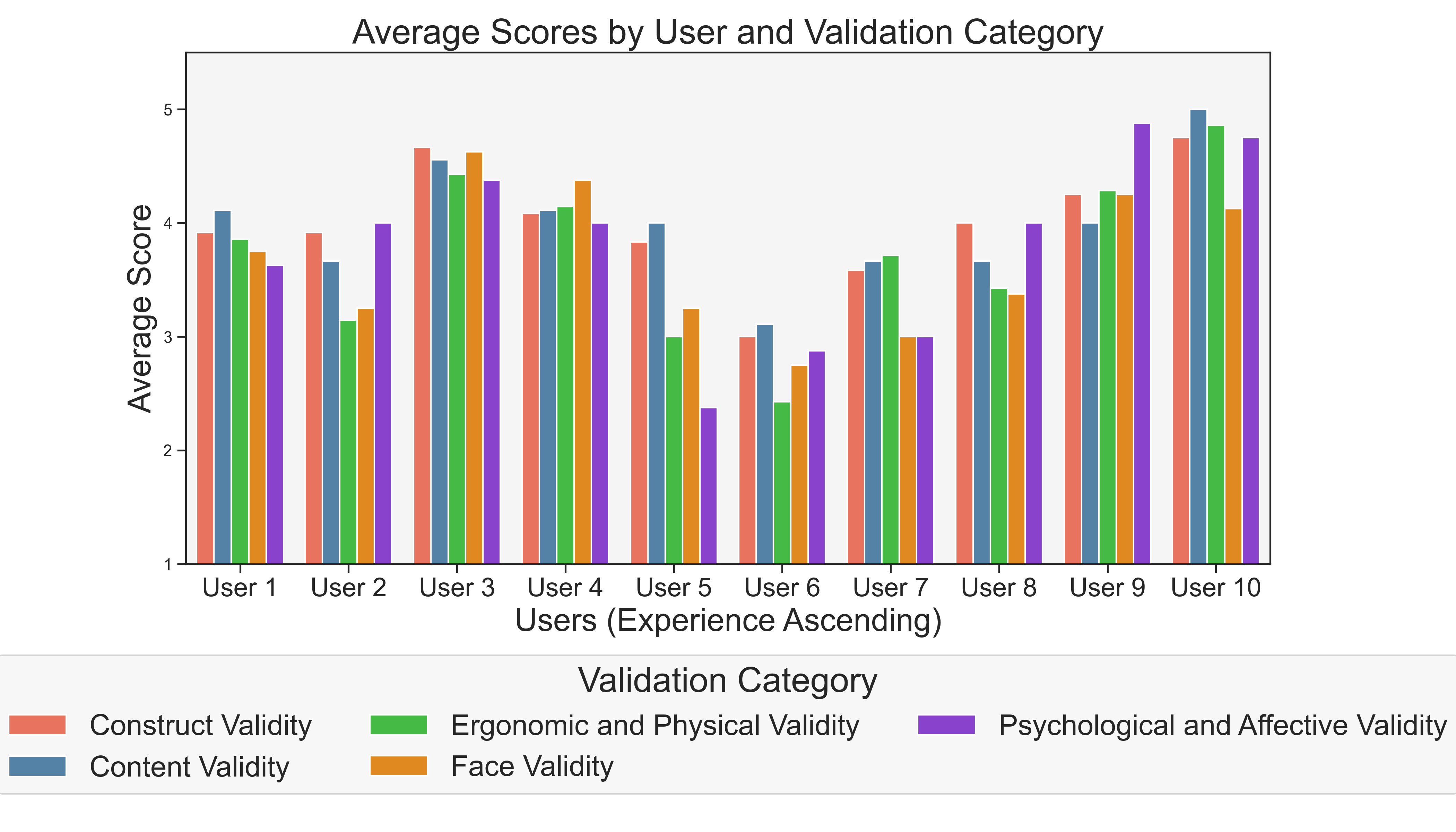}
    \caption{Average User Score Per Validation Category}
    \label{fig:validity_avg}
  \end{subfigure}
  \hfill
  \begin{subfigure}{0.49\textwidth}
    \centering
    \includegraphics[width=\linewidth]{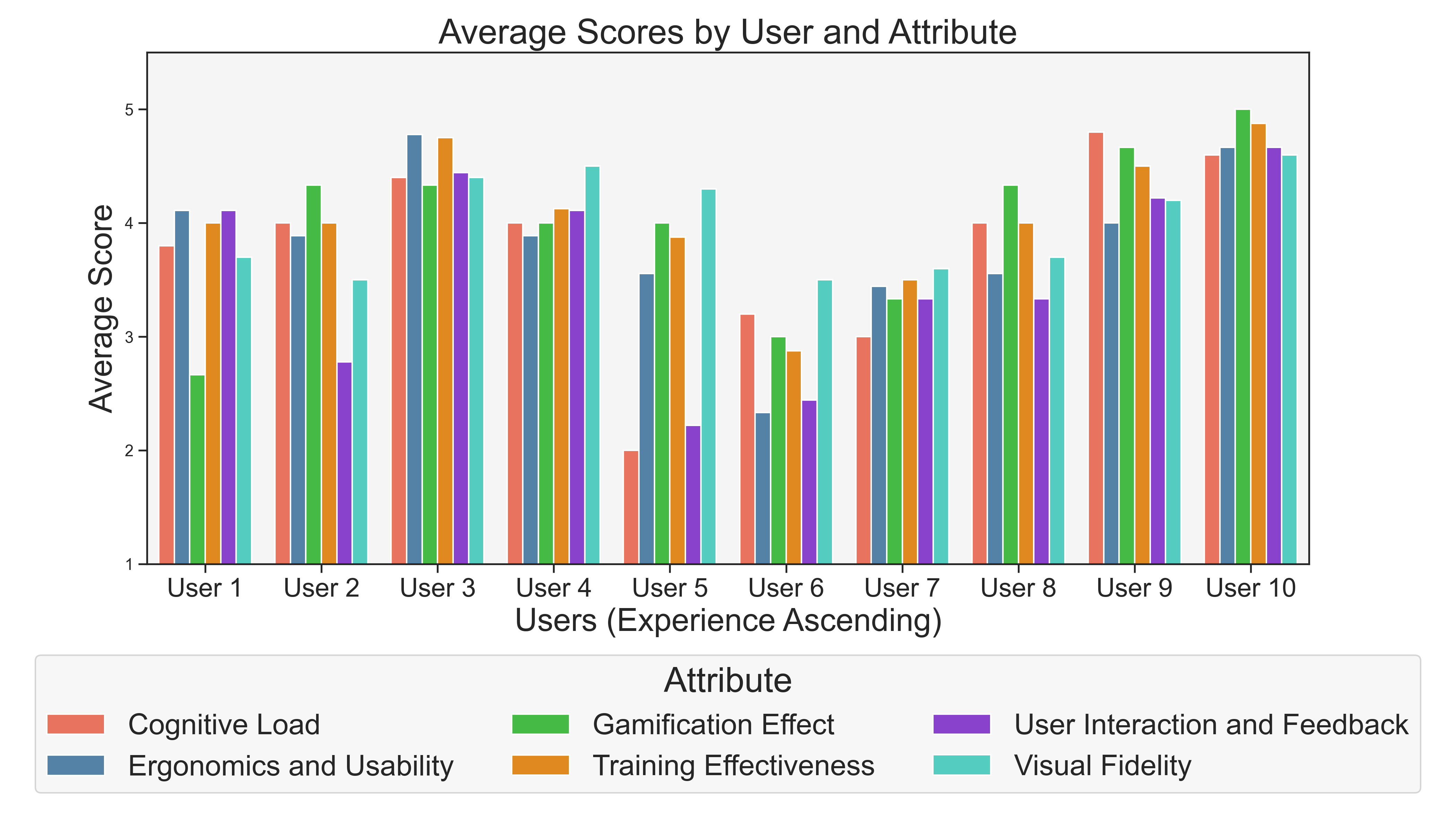}
    \caption{Average User Score Per Attribute}
    \label{fig:attribute_avg}
  \end{subfigure}
  
  \caption{Average user scores for different evaluation criteria}
  \Description{Two images arranged side by side, each with its own caption.}
\end{figure}

\begin{figure}[h]
  \centering
  \begin{subfigure}{0.49\textwidth}
    \centering
    \includegraphics[width=\linewidth]{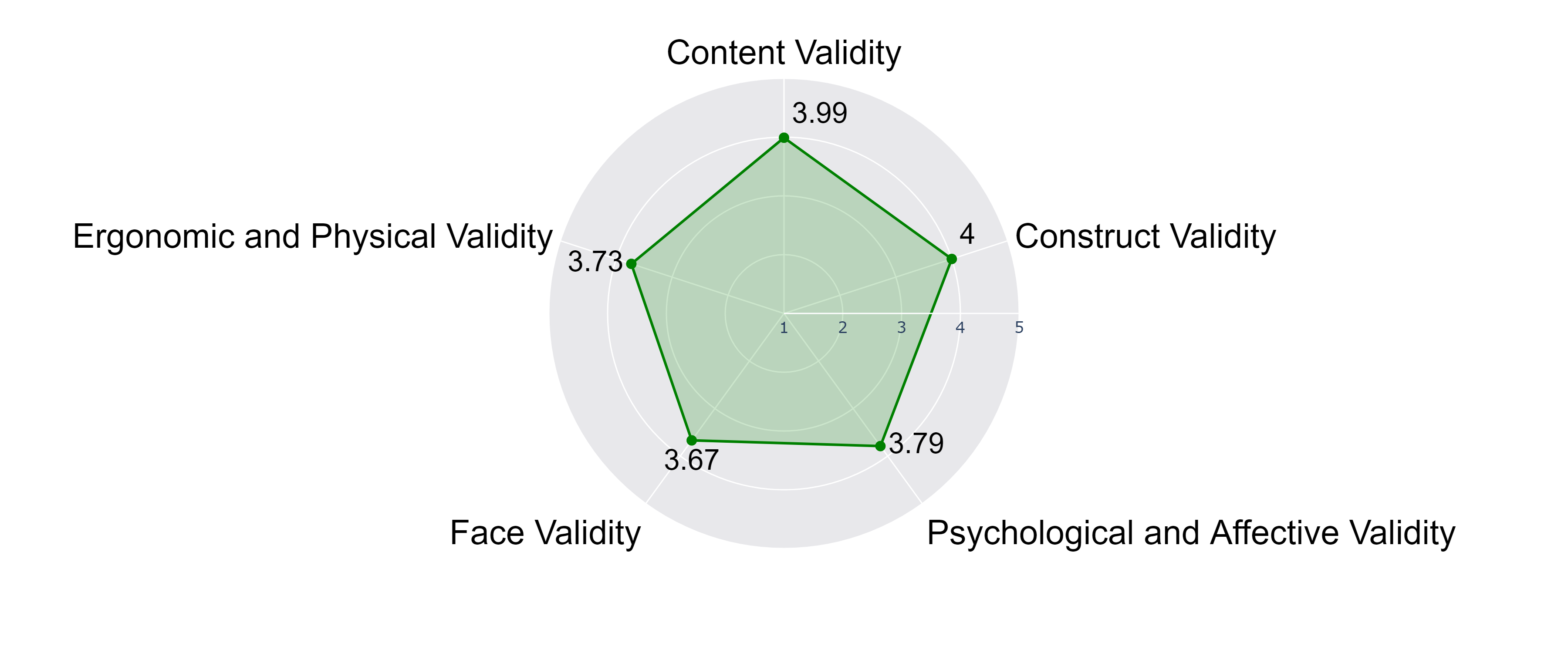}
    \caption{Average User Score Per Validation Category}
    \label{fig:validity_spider}
  \end{subfigure}
  \hfill
  \begin{subfigure}{0.49\textwidth}
    \centering
    \includegraphics[width=\linewidth]{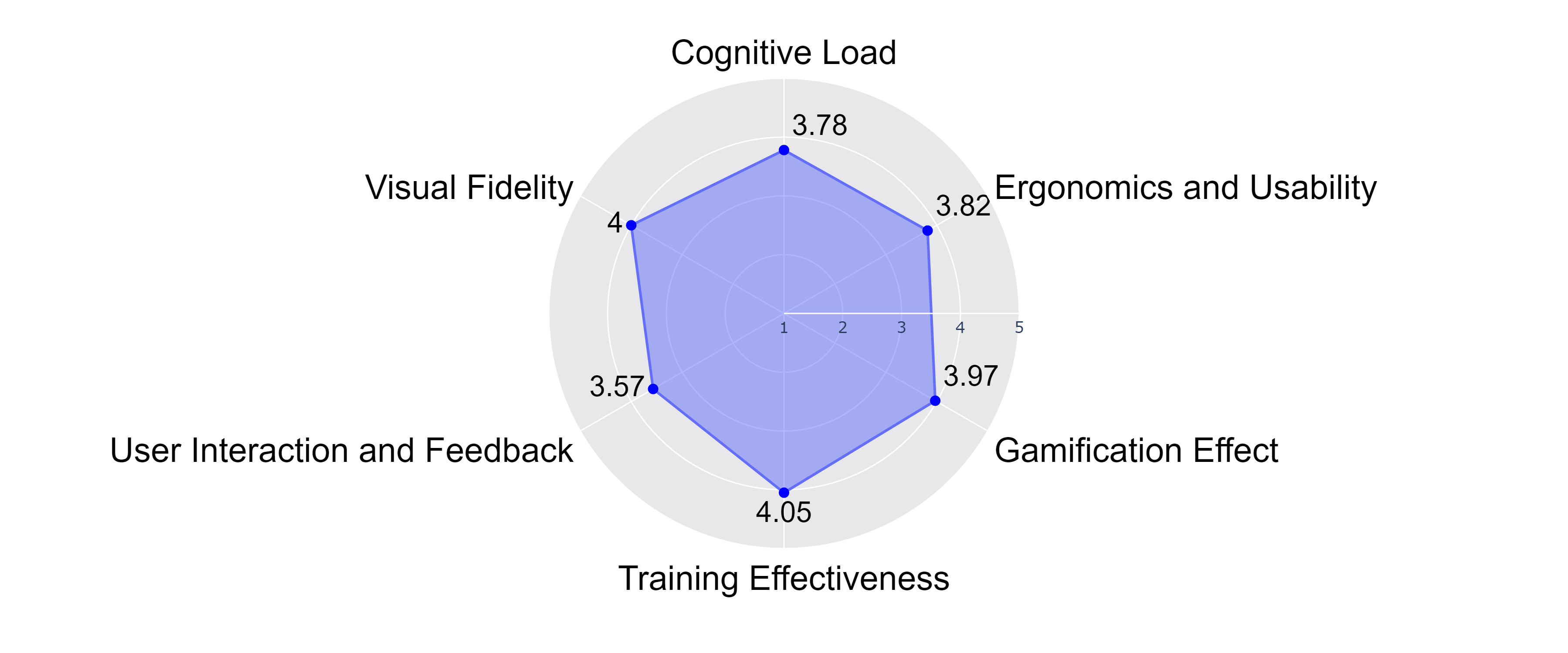}
    \caption{Average User Score Per Attribute}
    \label{fig:attribute_spider}
  \end{subfigure}
  
  \caption{Average user scores per a) validation category ,b) attribute}
  \Description{Two images arranged side by side, each with its own caption.}
\end{figure}

\subsection{Observations}
Both psychological \& affective validity and construct validity achieved moderate scores, averaging around 4.0, suggesting adequate cognitive and emotional engagement. However, the lower ergonomic and physical validity emphasizes the importance of ergonomic enhancements to improve user comfort and reduce cognitive strain. The results indicate that EchoSim4D performs well in terms of visual fidelity and engagement but could benefit from ergonomic improvements to enhance its effectiveness as a training tool.

\section{Limitations}
Manikin Setup and Sensor Placement: The current manikin setup serves as a proof of concept, with sensors affixed using glue. This temporary method highlights the need for more robust structures to ensure sensors remain static in precise locations. Future iterations will incorporate secure mounting solutions to enhance accuracy and reliability.
Hardware Performance: The simulator’s responsiveness is limited by the processing capabilities of the Arduino Nano and the latency of the MPU6050 sensor. Upgrading to more powerful microcontrollers and sensors with lower latency could improve overall performance, particularly for high-frequency applications.
VTK Plugin Optimization: The VTK plugin requires further optimization to perform smoothly on low-end systems. This refinement would enhance the accessibility of the simulator across a broader range of hardware configurations.
Mapping Precision: Delays in transitioning between apical normal and tilt positions indicate a need for more precise data mapping on the tilt axis. Improving this transition could provide smoother maneuvering and greater fidelity.
Magnetic Interference: External magnetic fields impact sensor accuracy, underscoring the need for shielding or alternative, interference-resistant sensor technologies in future versions.
Sensor Latency and Calibration: Current sensor latency limits real-time feedback, and the 20-second calibration duration disrupts workflow. Adopting faster sensors and optimizing calibration procedures could streamline the user experience.
Angle Tolerance and Ergonomics: The ±5° error margin may limit assessment precision, while ergonomic enhancements would reduce user fatigue and simulate clinical settings more accurately.

\section{Discussion \& Conclusion}
\textit{Overview of Findings}: EchoSim4D provides an effective training platform for neonatal echocardiography using a gamified, 4D visualization approach. Although successful as a proof of concept, addressing manikin setup and sensor placement issues will be crucial for more accurate simulations. Optimizing the VTK plugin will also improve accessibility, especially for users with low-end systems.
Enhanced Gamification and Practical Implications Additional gamification elements could improve user engagement and educational outcomes. For example, incorporating more complex interactive tasks could make the learning experience more immersive. In addition, a more accurate manikin setup and improved sensor placement would directly contribute to the fidelity of clinical skill acquisition.
User Feedback and Realism Enhancements: The temporary sensor affixation method impacted the consistency of feedback, underscoring the importance of stable sensor placement. User feedback also pointed to the need for tactile enhancements, particularly in terms of quality of manikin material, to better reflect the clinical experience.

\textit{Conclusion} EchoSim4D introduces a novel, accessible approach to training for neonatal echocardiography, utilizing a proof-of-concept manikin setup and real-time visualization techniques. While the current model demonstrates feasibility, future improvements in sensor placement and manikin design will enhance the realism and reliability of the simulator. Future work will focus on developing a refined manikin with stable sensor structures, enabling accurate sensor positioning. Additionally, plans include collecting data beyond standard TTE locations to comprehensively map the torso of a neonate, which will support a more accurate and immersive simulation experience. Expanding the simulator to cover diverse 4D ultrasound applications will further broaden its relevance and utility in medical training.

\bibliographystyle{ACM-Reference-Format}
\bibliography{bibliography}


\begin{thebibliography}{9}


\ifx \showCODEN    \undefined \def \showCODEN     #1{\unskip}     \fi
\ifx \showDOI      \undefined \def \showDOI       #1{#1}\fi
\ifx \showISBNx    \undefined \def \showISBNx     #1{\unskip}     \fi
\ifx \showISBNxiii \undefined \def \showISBNxiii  #1{\unskip}     \fi
\ifx \showISSN     \undefined \def \showISSN      #1{\unskip}     \fi
\ifx \showLCCN     \undefined \def \showLCCN      #1{\unskip}     \fi
\ifx \shownote     \undefined \def \shownote      #1{#1}          \fi
\ifx \showarticletitle \undefined \def \showarticletitle #1{#1}   \fi
\ifx \showURL      \undefined \def \showURL       {\relax}        \fi
\providecommand\bibfield[2]{#2}
\providecommand\bibinfo[2]{#2}
\providecommand\natexlab[1]{#1}
\providecommand\showeprint[2][]{arXiv:#2}

\bibitem[Allgaier et~al\mbox{.}(2024)]%
        {gamificliver}
\bibfield{author}{\bibinfo{person}{Mareen Allgaier}, \bibinfo{person}{Florentine Huettl}, \bibinfo{person}{Laura~Isabel Hanke}, \bibinfo{person}{Tobias Huber}, \bibinfo{person}{Bernhard Preim}, \bibinfo{person}{Sylvia Saalfeld}, {and} \bibinfo{person}{Christian Hansen}.} \bibinfo{year}{2024}\natexlab{}.
\newblock \showarticletitle{Gamification Concepts for a VR-based Visuospatial Training for Intraoperative Liver Ultrasound}. In \bibinfo{booktitle}{\emph{Extended Abstracts of the CHI Conference on Human Factors in Computing Systems}} (Honolulu, HI, USA) \emph{(\bibinfo{series}{CHI EA '24})}. \bibinfo{publisher}{Association for Computing Machinery}, \bibinfo{address}{New York, NY, USA}, Article \bibinfo{articleno}{175}, \bibinfo{numpages}{8}~pages.
\newblock
\showISBNx{9798400703317}
\urldef\tempurl%
\url{https://doi.org/10.1145/3613905.3650736}
\showDOI{\tempurl}


\bibitem[{Elevate Health Inc.}(2024)]%
        {vimedix}
\bibfield{author}{\bibinfo{person}{{Elevate Health Inc.}}} \bibinfo{year}{2024}\natexlab{}.
\newblock \bibinfo{booktitle}{\emph{Vimedix}}.
\newblock
\urldef\tempurl%
\url{https://elevatehealth.net/solutions/brands/vimedix}
\showURL{%
\tempurl}


\bibitem[{Epic Games}(2024)]%
        {unrealengine}
\bibfield{author}{\bibinfo{person}{{Epic Games}}.} \bibinfo{year}{2024}\natexlab{}.
\newblock \bibinfo{booktitle}{\emph{Unreal Engine}}.
\newblock
\urldef\tempurl%
\url{https://www.unrealengine.com}
\showURL{%
\tempurl}


\bibitem[{Intelligent Ultrasound}(2024)]%
        {heartworks}
\bibfield{author}{\bibinfo{person}{{Intelligent Ultrasound}}.} \bibinfo{year}{2024}\natexlab{}.
\newblock \bibinfo{booktitle}{\emph{Heartworks}}.
\newblock
\urldef\tempurl%
\url{https://www.intelligentultrasound.com/heartworks/}
\showURL{%
\tempurl}


\bibitem[Krishnamurthy et~al\mbox{.}(2024)]%
        {gamificationmeded}
\bibfield{author}{\bibinfo{person}{Kandamaran Krishnamurthy}, \bibinfo{person}{Nikil Selvaraj}, \bibinfo{person}{Palak Gupta}, \bibinfo{person}{Benitta Cyriac}, \bibinfo{person}{Puvin Dhurairaj}, \bibinfo{person}{Adnan Abdullah}, \bibinfo{person}{Ambigga Krishnapillai}, \bibinfo{person}{Halyna Lugova}, \bibinfo{person}{Mainul Haque}, \bibinfo{person}{Sophie Xie}, {and} \bibinfo{person}{Eng-Tat Ang}.} \bibinfo{year}{2024}\natexlab{}.
\newblock \showarticletitle{Benefits of gamification in medical education}.
\newblock \bibinfo{journal}{\emph{Clinical Anatomy}} (\bibinfo{year}{2024}).
\newblock
\urldef\tempurl%
\url{https://doi.org/10.1002/ca.23916}
\showDOI{\tempurl}


\bibitem[Lee(2020)]%
        {Unity}
\bibfield{author}{\bibinfo{person}{Newton Lee}.} \bibinfo{year}{2020}\natexlab{}.
\newblock \bibinfo{booktitle}{\emph{Unity, A 2D and 3D Game Engine}}.
\newblock \bibinfo{publisher}{Springer International Publishing}, \bibinfo{address}{Cham}, \bibinfo{pages}{1--3}.
\newblock
\showISBNx{978-3-319-08234-9}
\urldef\tempurl%
\url{https://doi.org/10.1007/978-3-319-08234-9_536-1}
\showDOI{\tempurl}


\bibitem[Nicholls et~al\mbox{.}(2014)]%
        {psychomotor}
\bibfield{author}{\bibinfo{person}{D. Nicholls}, \bibinfo{person}{L. Sweet}, {and} \bibinfo{person}{J. Hyett}.} \bibinfo{year}{2014}\natexlab{}.
\newblock \showarticletitle{Psychomotor skills in medical ultrasound imaging}.
\newblock \bibinfo{journal}{\emph{Journal of Ultrasound in Medicine}}  \bibinfo{volume}{33} (\bibinfo{year}{2014}), \bibinfo{pages}{1349--1352}.
\newblock
Issue 8.
\urldef\tempurl%
\url{https://doi.org/10.7863/ultra.33.8.1349}
\showDOI{\tempurl}


\bibitem[Schroeder et~al\mbox{.}(2006)]%
        {vtkBook}
\bibfield{author}{\bibinfo{person}{Will Schroeder}, \bibinfo{person}{Ken Martin}, {and} \bibinfo{person}{Bill Lorensen}.} \bibinfo{year}{2006}\natexlab{}.
\newblock \bibinfo{booktitle}{\emph{The Visualization Toolkit (4th ed.)}}.
\newblock \bibinfo{publisher}{Kitware}.
\newblock
\showISBNx{978-1-930934-19-1}


\bibitem[{SonoSim, Inc.}(2024)]%
        {sonosim}
\bibfield{author}{\bibinfo{person}{{SonoSim, Inc.}}} \bibinfo{year}{2024}\natexlab{}.
\newblock \bibinfo{booktitle}{\emph{SonoSim}}.
\newblock
\urldef\tempurl%
\url{https://sonosim.com/}
\showURL{%
\tempurl}


\end{thebibliography}

\appendix
\section{Information About TTE Views}

EchoSim4D incorporates these views to simulate neonatal cardiac assessment:

\begin{enumerate}
    \item \textbf{Apical Views (Normal and Tilt)}: Essential for evaluating ventricles, atria, and valves. The Tilt view provides enhanced visualization of the left ventricular outflow tract (LVOT), allowing detailed insights into ventricular dynamics.
    \item \textbf{Parasternal Short-Axis (PSAX) Views (Normal and Tilt)}: These views offer a comprehensive look at left ventricular function, morphology, and septal motion. The Tilt view aids in diagnosing congenital heart defects by adding perspectives on LVOT and outflow tracts.
    \item \textbf{Parasternal Long-Axis (PLAX) Views (Normal and Tilt)}: Foundational for visualizing the left atrium, left ventricle, mitral valve, aortic valve, and LVOT, these views are critical for assessing ventricular outflow.
    \item \textbf{Subcostal Views (Normal and Tilt)}: Particularly valuable for neonates, allowing imaging through the liver to evaluate the atrial septum and vena cava. The Tilt view helps identify shunts and anomalies affecting the LVOT.
    \item \textbf{Suprasternal Views (Normal and Tilt)}: Key for assessing the aortic arch and great vessels, with the Tilt view enhancing vascular visualization. These views are more focused on vascular structures than on ventricular or atrial visualization.
\end{enumerate}

\end{document}